\documentclass[showpacs,preprintnumbers,twocolumn,superscriptaddress]{revtex4-1}
\usepackage{bm, mathrsfs, amsmath, amssymb, here}
\usepackage[dvipdfmx]{graphicx}

\begin{document}
\title{Emergence of soft quark excitations by the coupling \\
with a soft mode of the QCD critical point}

\author{Masakiyo Kitazawa}
\affiliation{Department of Physics, Osaka University, Toyonaka,
Osaka 560-0043, Japan}
\author{Teiji Kunihiro}
\affiliation{Department of Physics,
Kyoto University, Kyoto 606-8502, Japan}
\author{Yukio Nemoto}
\affiliation{Department of Physiology, 
St. Marianna University School of Medicine,
Kanagawa 216-8511, Japan}

\begin{abstract}
We study the quark spectrum at nonzero temperature and density
near the critical point (CP) of the chiral phase transition 
incorporating effects of the scalar- and pseudoscalar-density
fluctuations in a chiral effective model with a nonzero current quark mass.
It is known that the soft mode associated with the second-order transition at the CP lies 
in the spacelike region of the scalar-density fluctuation.
We find that the soft mode influences the quark spectrum
significantly near the CP, resulting in the shift of the quasiquark peak.
Effects of the composite stable pions on the quark spectrum near the CP are also discussed.
\end{abstract}
\date{\today}

\preprint{KUNS-2517}

\maketitle

\section{Introduction} \label{sec:intro}

The phases of quantum chromodynamics (QCD) at nonzero temperature ($T$) and
the quark-chemical potential ($\mu$) are of much interest both theoretically 
and experimentally.
It has been established in lattice QCD with 
the physical quark masses that there does not
exist a genuine phase transition at nonzero temperature and zero baryon density
but only a rapid crossover from the hadronic state to the quark-gluon plasma 
\cite{Bernard:2004je, Cheng:2006qk, Aoki:2006we, Ejiri:2009ac}.
In the cold and dense region $\mu \gtrsim T$, in contrast, reliable lattice simulations  
are still missing because of the notorious sign problem, 
although  various attempts have been made
\cite{Fodor:2004nz,Gavai:2004sd,Philipsen:2011zx,Ejiri:2013lia,Aarts:2013uxa}.
Low-energy effective models of QCD 
\cite{Klevansky:1992qe,Hatsuda:1994pi,Buballa:2003qv} tell us
that there is a critical point (CP) 
at $T>0$ and $\mu>0$ where the critical line of the first-order phase transition in the 
low-temperature region terminates and that becomes second order
\cite{Asakawa:1989bq}.
We mention that there can be variants or alternatives of the phase diagram with 
multiple or no critical points 
when the vector interaction \cite{Kunihiro:1991qu}, color superconductivity with 
and without  charge neutrality and/or the anomaly terms are incorporated \cite{Kitazawa:2002bc,Zhang:2008wx,Hatsuda:2006ps,deForcrand:2006pv,Basler:2010xy,Zhang:2011xi},
even apart from the possibility of inhomogeneous phases
\cite{Nakano:2004cd,Nickel:2009ke}. 

The global structure of the QCD phase diagram, including
the identification of the CP,
can be investigated experimentally 
using relativistic heavy ion collisions by comparing 
the collision events with different collision energies.
An experimental program to perform such an investigation,
called the beam energy scan program, is now ongoing 
at the Relativistic Heavy Ion Collider \cite{Aggarwal:2010wy}.
The existence of the CP may be confirmed by the measurement
of fluctuation observables in this program \cite{Stephanov:1998dy}.
Experiments at the 
planned facilities at  GSI, NICA and J-PARC
would also contribute to reveal  
physical properties of the CP as well as the its existence.

We emphasize here that 
the CP  appears when the
chiral symmetry is explicitly broken, 
and the identifications of the order parameter of the second-order transition 
and the associated soft modes 
are {\em not} simple and involved \cite{Fujii:2003bz,Son:2004iv}.
In the chiral limit where the current $u, d$ quarks are massless, the order parameter {\em is}
the chiral condensate, and 
the critical line of the first-order transition in the low-temperature region is 
 connected at the {\em tricritical point} (TCP) 
with the critical line of the second-order transition; the 
associated soft modes to the phase transition
are the amplitude and phase fluctuations of the order parameter, which are identified
with the sigma and the pionic modes, respectively \cite{Hatsuda:1984jm}. We notice that 
there is no scalar-vector coupling at the TCP due to the chiral symmetry.

When  the chiral symmetry is explicitly broken with the nonzero current quark masses,
the second-order transition in the high-temperature region turns to a crossover and the sigma 
and pionic modes stay massive even at the CP: Indeed the sigma mode 
has a mass $m_\sigma \simeq 2m$ around the CP with
 $m$ being the dynamically generated constituent quark mass.
Moreover, in such a case with an explicit chiral-symmetry breaking at $\mu\not=0$,
the scalar-vector cross-correlation term $\langle :(\bar{\psi}\gamma^0\psi)(\bar{\psi}\psi):\rangle$  
does not vanish, and, hence, the scalar mode  is coupled with
the density-density correlator $\langle :(\bar{\psi}\gamma^0\psi)^2:\rangle$ \cite{Kunihiro:1991qu};
this is because charge conjugation symmetry is violated with nonzero $\mu$,
and the left- and right-handed quarks are coupled owing to the breaking
of chiral symmetry. It has been shown \cite{Fujii:2003bz,Son:2004iv} 
that the soft mode, the mass of which {\em vanishes} at the CP, is a superposition of the 
number-density fluctuation  (phonon) and the sigma mode, and 
the QCD CP belongs to the same universality class as the liquid-gas phase 
transition, which is  called  ``model H" in the classification scheme by Hohenberg and Halperin \cite{Hohenberg:1977ym}:
Precisely speaking, the soft mode associated with the CP consists of not only the 
sigma and the phonon mode but also the entropy fluctuations, i.e., the hydrodynamic 
modes \cite{Fujii:2003bz,Son:2004iv}.

In the present work, 
we investigate how the  soft mode associated with the CP 
affects in turn the quark spectrum near the CP.
We employ the simplest Nambu--Jona-Lasinio (NJL) model \cite{Nambu:1961tp,Klevansky:1992qe,Hatsuda:1994pi,Buballa:2003qv}
without the vector coupling to describe the
soft mode associated with the CP.
In this model, the collective mode in the scalar channel has a support 
in the spectral function not only in the timelike region 
but also in the spacelike region, the latter of 
which is composed of  particle-hole excitations and describes the soft mode 
associated with the CP \cite{Fujii:2003bz}.

It is known that the quarks coupled with the bosonic excitations 
at nonzero temperature show an unexpected rich structure in the spectral function,
even apart from the formation of the celebrated plasmino excitations at extremely high temperatures
as given in the hard-thermal-loop (HTL) approximation in gauge theories \cite{Frenkel:1989br,LeB}.
In \cite{Kitazawa:2005mp},   the  spectral properties of quarks coupled with the soft modes 
at zero density were examined  above but near the critical temperature
of the second-order chiral transition in the chiral limit.
It was found that a novel peak is formed around the zero energy as well as
the normal and the antiplasmino ones, which then makes 
a three-peak structure in the quark spectral function:
The formation of the far-low-lying peak is owed
to the mixing between a quark (antiquark) and an antiquark hole 
(quark hole) by a resonant scattering of the quasiquarks with the soft modes with small but
nonzero masses \cite{Kitazawa:2005mp, Kitazawa:2006zi, Satow:2010ia}.
Such a formation of the three-peak structure in the spectral function of a fermion coupled with 
a bosonic excitation  at $T\not=0$
is now confirmed beyond one loop \cite{Harada:2007gg, Harada:2008vk, Qin:2010pc,
Nakkagawa:2011ci, Qin:2013ufa}.
Furthermore, a similar third peak in the ultrasoft region 
has been shown to exist in gauge theories at high temperature beyond the 
HTL approximation
on the basis of a novel resummation technique \cite{Hidaka:2011rz}.
It is also shown that the neutrino spectrum shows 
similar three peaks containing an ultrasoft branch at 
electroweak scale temperature \cite{Miura:2013fxa}.

It should be noted, however,  that  
such a three-peak structure in the fermion spectrum tends to be  suppressed
as the fermion mass becomes large \cite{Kitazawa:2007ep}.
Thus one might naturally suspect that
the quark spectrum near the CP, which is realized when the chiral symmetry is
explicitly broken,  would not have any anomalous structure 
like that seen near the critical temperature in the chiral limit.
However, a recent paper \cite{KKN} of 
the present authors shows that it may not be the case:
In fact,  the quark spectral function in this region can have an 
anomalously low-lying peak
due to the coupling with the stable pionic modes with a nonhyperbolic dispersion relation
around the pseudocritical temperature, 
which leads to a van Hove singularity of the joint density of states in the imaginary part of 
the quark self energy.

In the present paper,  
we shall show that the quark spectrum around the CP
shows yet another unexpected behavior due to the coupling 
to the phononlike soft mode in the spacelike region.
The spectral properties of fermions at nonzero $\mu$ have been
studied in gauge theories in the hard dense limit \cite{LeB}
and also the electron spectrum in high density plasma \cite{HL}.
The modification of the quark spectrum near the CP is different
from both of these cases, because the soft mode associated with the CP 
consists of particle-hole states in the spacelike region.
It will also be shown that the effects of the van Hove singularity 
present around the crossover region at $\mu=0$
almost die out around the CP.

The paper is organized as follows.
In the next section, we introduce the NJL model to investigate 
the phase diagram of the chiral transition, 
and formulate the fluctuation modes and the quark self energy.
We then discuss the properties of the fluctuation modes in 
Sec.~\ref{sec:softmode}.
The numerical results are shown in Sec.~\ref{sec:num}.
A summary and concluding remarks are given in Sec.~\ref{sec:conc}.

\section{Formalism}
\label{sec:pd}

In this section, after introducing the model used in this study
we present the phase diagram obtained in this model.
We then give the expressions of 
the meson and quark propagators in the random-phase approximation at nonzero $T$ and $\mu$.
The calculation procedures described in Secs.~\ref{sec:sigma}$\sim$\ref{sec:selfe}
 are basically the same as those in our 
previous study \cite{KKN}, except for the 
introduction of nonzero quark chemical potential $\mu$.

\subsection{Model and phase diagram}
\label{sec:model}

We employ the two-flavor NJL model
\cite{Nambu:1961tp}
\begin{equation}
  \mathcal{L}=\bar{\psi} (i \partial \hspace{-0.5em} /  - m_0) \psi
  + G_S [(\bar{\psi} \psi)^2 + (\bar{\psi}i\gamma_5{\boldsymbol \tau}\psi)^2],
  \label{eq:NJL}
\end{equation}
as an effective model of low-energy QCD \cite{Hatsuda:1994pi},
and study the effect of fluctuation modes in the scalar $(\sigma)$ and
pseudoscalar $(\pi)$ channels on the spectral properties of
quarks near the phase boundary,
with $\boldsymbol \tau$ being the flavor SU(2) Pauli matrices.
The coupling constant $G_S=5.5$ GeV${}^{-2}$, the current $u,d$-quark
mass $m_0=5.5$ MeV, and the three-dimensional
cutoff $\Lambda=631$ MeV are determined so as to reproduce 
the pion mass, the pion decay constant, and the quark condensate 
in vacuum \cite{Hatsuda:1994pi}.

\begin{figure}[t]
\includegraphics[width=8cm]{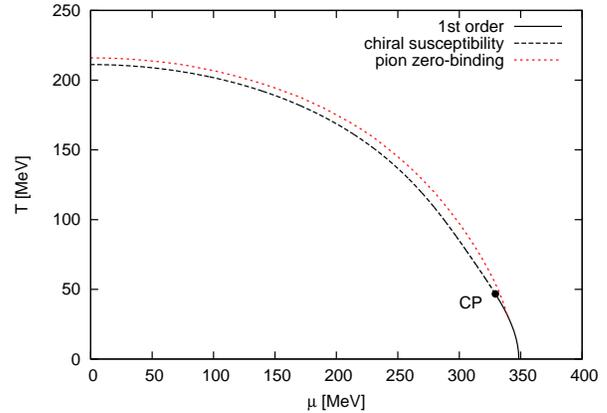}
\caption{The phase diagram of the chiral transition.
CP means the critical point.
The solid line denotes the first-order phase transition,
the dashed line the maximum of the chiral susceptibility at each $\mu$.
The dotted line denotes the pion zero-binding temperatures where
the binding energy of the stable pionic modes vanishes.}
\label{fig:phased}
\end{figure}

As mentioned in Sec.~\ref{sec:intro},
a complete description of the effects of the specific collective
modes on the quark spectrum necessitates the inclusion of the
couplings of the number-density and the entropy fluctuations
to the quark field.
There are some attempts to do it:
A model Lagrangian has been proposed to take into
account the phonon-quark coupling in the meson-quark model
\cite{Kamikado:2012cp}.
We also note that although in a quite different context,
Shen and Reddy \cite{Shen:2013kxa} calculated  the neutrino 
scattering from hydrodynamic
modes in hot and dense neutron matter  using 
a transport equation. However, an extension of the model for  such a complete description 
of the hydrodynamic modes and their couplings to the quark is beyond the scope of the present work.

In Fig.~\ref{fig:phased}, we show the phase diagram of the chiral
transition 
in the mean-field approximation (MFA).
There is a first-order transition at low temperature and large 
chemical potential region 
shown by the solid line, which terminates at the CP,
denoted by CP in the figure,
with the critical temperature $T_c$ and the chemical potential $\mu_c$ 
given by
\[
T_c\simeq47 {\rm MeV},\quad  
\mu_c\simeq329 {\rm MeV}.
\]
The phase transition is of second order at this point.
The transition becomes crossover for $\mu<\mu_c$.
For a guide of the phase boundary for $\mu<\mu_c$,
the temperatures at which the chiral susceptibility has the 
maximum at a fixed chemical potential, i.e., the ridge of the 
chiral susceptibility, are shown by the dashed line.

\subsection{Fluctuation modes}
\label{sec:sigma}

Next we construct the propagator of 
the scalar and pseudoscalar fluctuations at nonzero $T$ and $\mu$.
The soft mode associated with the CP manifests itself in the
scalar fluctuations.
The pions, which are described by the pseudoscalar fluctuations, 
are still bound
states near the CP in this model as shown 
in Sec.~\ref{sec:softmode}.
The excitation properties of these fluctuation modes are characterized by 
the spectral functions for the quark-antiquark excitations
in the $\sigma$ and $\pi$ channels
\begin{equation}
  \rho_{\sigma(\pi)} ( \bm{p},p_0 )
  = -\frac1\pi {\rm Im} D_{\sigma(\pi)}^R ( \bm{p},p_0 ) ,
 \label{eq:spc-sp}
\end{equation}
where $D_\sigma^R (\bm{p},p_0)$ and $D_\pi^R (\bm{p},p_0)$
are the retarded quark-antiquark propagators in each channel.
The corresponding imaginary-time propagators in the 
random-phase approximation read
\begin{equation}
  {\cal D}_{\sigma(\pi)}(\bm{p},\nu_n) 
  = -\frac{1}{1/(2G_S)+{\cal Q}_{\sigma(\pi)}(\bm{p},\nu_n)} ,
  \label{eq:d-sp}
\end{equation}
where ${\cal Q}_\sigma(\bm{p},\nu_n)$ and ${\cal Q}_\pi(\bm{p},\nu_n)$ 
are the one-loop quark-antiquark polarization functions
\begin{align}
  {\cal Q}_{\sigma}(\bm{p},\nu_n) &= T\sum_m \int \frac{d^3q}{(2\pi)^3}
  {\rm Tr}[{\cal G}_0(\bm{q},\omega_m) \notag \\
  &\ \times{\cal G}_0(\bm{p}+\bm{q},\nu_n+\omega_m)], 
  \label{eq:Q-sig}  \\
  {\cal Q}_{\pi}(\bm{p},\nu_n) &= \frac{T}{3}\sum_m \int \frac{d^3q}{(2\pi)^3}
  {\rm Tr}[i\gamma_5 \bm{\tau} 
  {\cal G}_0(\bm{q},\omega_m) \notag \\
  &\ \times i\gamma_5\bm{\tau}{\cal G}_0
  (\bm{p}+\bm{q},\nu_n+\omega_m)],
  \label{eq:Q-pi}
\end{align}
with $\mathcal{G}_0(\bm{p},\omega_n)=[(i\omega_n+\mu) \gamma_0
-\bm{p}\cdot\vec{\gamma}-m]^{-1}$ being the 
quark propagator
and $\nu_n=2n\pi T$ and $\omega_n=(2n+1)\pi T$ denoting 
the Matsubara frequencies for bosons and fermions, respectively.
$m=m_0+m_D$ is the constituent quark mass where $m_D$ is the
dynamically generated mass from spontaneous chiral-symmetry breaking
evaluated in the MFA.
After the summation of the Matsubara frequency and the analytic 
continuation with a replacement $i\nu_n\to p_0+i\eta$, 
we obtain the retarded functions $Q_{\sigma(\pi)}^R ( \bm{p},p_0 )$ and
 $D_{\sigma(\pi)}^R ( \bm{p},p_0 )$.
For the numerical calculation of $Q^R_{\sigma(\pi)}( \bm{p},p_0 )$, 
we first calculate the imaginary part 
that is free from the ultraviolet divergence
and then evaluate
the real part with the Kramers-Kronig relation
\begin{equation}
  {\rm Re}Q^R_{\sigma(\pi)}(\bm{p},p_0) = 
  -\frac{1}{\pi}{\rm P}\int_{-\Lambda'}^{\Lambda'} dp_0' 
  \frac{{\rm Im}Q^R_{\sigma(\pi)}(\bm{p},p_0')}{p_0-p_0'} ,
\label{eq:KramersKronig}
\end{equation}
where P denotes the principal value and the energy cutoff 
$\Lambda'=2\sqrt{\Lambda^2+m^2}$ is determined such that 
$D_\sigma^R(\bm{p},p_0)$ 
at $\bm{p}=0$ and $p_0=0$ in the spacelike region
diverges at the CP determined by the MFA.
This is a softening condition associated with the CP
and is consistent with the introduction of the cutoff $\Lambda$
into the thermodynamic potential in the MFA.

The imaginary parts of $Q^R_{\sigma}(\bm{p},p_0)$ and
$Q^R_{\pi}(\bm{p},p_0)$ are proportional to the difference 
between the decay and creation rates of each mode
and take nonzero values for $ |p_0| > E_{\rm thr}(p)$
and $ |p_0| < p $, with $E_{\rm thr}(p)=\sqrt{ p^2 + 4m^2 }$
and $p=|\bm{p}|$.
The decay processes into a quark and an antiquark take place
for $ p_0 > E_{\rm thr}(p)$, 
where the threshold energy $E_{\rm thr}(p)$ 
is the lowest value of the sum of excitation energy of 
a quark and an antiquark  with a fixed total momentum $p$.
The decay process in the spacelike region, $|p_0|<p$, is
the Landau damping, which is the scattering processes of a
quark or an antiquark with the fluctuation modes.

The propagators $D_{\sigma(\pi)}^R(\bm{p},p_0)$ may have poles on the
lower-half complex energy plane corresponding to 
collective excitations.
In vacuum, 
$D_{\pi}^R(\bm{p},p_0)$ has a bound-state pole on the real axis 
corresponding to the pseudo--Nambu-Goldstone pion \cite{KKN}.
The pole of $D_{\sigma}^R(\bm{p},p_0)$, on the other hand, 
has an imaginary part, which implies that the sigma meson 
is a resonance state with a decay width. 

When the poles of $D_{\pi}^R(\bm{p},p_0)$ are located on the real axis,
the dispersion relation of the stable pionic modes,
$\omega_\pi(p)$, is obtained by solving the equation
\begin{align}
  [{\rm Re}D_\pi^R(\bm{p},\omega_\pi(p))]^{-1}
  &=-\frac{1}{2G_S}-{\rm Re}Q_\pi^R(\bm{p},\omega_\pi(p)) \notag \\
  &=0.
  \label{omegapi}
\end{align}
Because ${\rm Im}D^R_\pi(\bm{p},p_0)$ vanishes for 
$p<|p_0|<E_{\rm thr}(p)$, Eq.~(\ref{omegapi}) 
gives the correct dispersion relation only for $p<\omega_\pi(p)<E_{\rm thr}(p)$.
The residue $Z_\pi(p)$ of the bound pole is given by
\begin{align}
  \frac{1}{Z_\pi(p)} &= -\frac{1}{\pi} \frac{\partial[D_\pi^R(\bm{p},\omega_\pi(p_0))]^{-1}}
  {\partial p_0} \bigg|_{p_0=\omega_\pi(p)} \notag \\
  &= -\frac{1}{\pi} \frac{\partial Q_\pi^R(\bm{p},\omega_\pi(p_0))}
  {\partial p_0} \bigg|_{p_0=\omega_\pi(p)} .
\end{align}
When Eq.~(\ref{omegapi}) has a solution in the range of $p_0$
at which ${\rm Im}D^R_\pi(\bm{p},p_0)$ takes a nonzero value,
we refer to the solution as the quasipole.

\subsection{Quark self energy}
\label{sec:selfe}

In this study, we consider the effect of the fluctuation modes
in the $\sigma$ and $\pi$ channels near the CP
on spectral properties of quarks.
As a first step of such an analysis, 
we use the random-phase approximation \cite{KKN},
where the fluctuation modes are 
constructed by the undressed quarks and antiquarks 
in a non--self-consistent way.
In addition, we limit our attention to 
the quark spectral function at zero momentum,
since the soft mode influences the quark spectrum at vanishing momentum
most strongly. 

The quark self energy in the random-phase approximation 
for $p=0$ is given by
\begin{align}
 &\ \tilde{\Sigma}(\bm{p}=\bm{0},\omega_n) \equiv \tilde{\Sigma}(\omega_n) \notag \\
  &= -T\sum_m \int\frac{d^3q}{(2\pi)^3} \big[
  \mathcal{D}_\sigma(\bm{q},\omega_n-\omega_m) \mathcal{G}_0(\bm{q},\omega_m) \notag \\
  &\quad\ +3\mathcal{D}_\pi(\bm{q},\omega_n-\omega_m) 
  i\gamma_5\mathcal{G}_0(\bm{q},\omega_m)i\gamma_5 \big] ,
  \label{eq:sgtld}
\end{align}
where the factor 3 in front of $\mathcal{D}_\pi$ comes from the isospin degeneracy.

After the summation of the Matsubara frequency in Eq.~(\ref{eq:sgtld})
and the analytic continuation, $i\omega_n\to p_0+i\eta$, we obtain
the retarded quark self energy,
\begin{equation}
  \Sigma^R(p_0) = \Sigma_\sigma^R(p_0)+\Sigma_\pi^R(p_0),
  \label{eq:sg}
\end{equation}
with
\begin{align}
  \Sigma_\sigma^R(p_0) &=  \int\frac{d^4q}{(2\pi)^4}
  \frac{(\gamma^0+m/E_q)\pi \rho_\sigma(\bm{q},q_0)}{q_0-p_0+E_q-\mu-i\eta} \notag\\
  &\ \times \left[  1+n(q_0) -f_-(E_q)  \right]
  \notag \\
  &  +\int\frac{d^4q}{(2\pi)^4} 
  \frac{(\gamma^0-m/E_q)\pi \rho_\sigma(\bm{q},q_0)}
  {q_0-p_0-E_q-\mu-i\eta} \notag \\
  &\ \times \left[  n(q_0) +f_+(E_q)  \right] ,
  \label{eq:sg1} 
\end{align}
and
\begin{align}
  \Sigma_\pi^R(p_0) &=  \int\frac{d^4q}{(2\pi)^4}
  \frac{(\gamma^0-m/E_q)3\pi \rho_\pi(\bm{q},q_0)}{q_0-p_0+E_q-\mu-i\eta}
  \notag \\
  &\ \times\left[  1+n(q_0) -f_-(E_q)  \right]
  \nonumber \\
  &  +\int\frac{d^4q}{(2\pi)^4}
  \frac{(\gamma^0+m/E_q)3\pi \rho_\pi(\bm{q},q_0)}
  {q_0-p_0-E_q-\mu-i\eta} \notag \\
  &\ \times \left[  n(q_0) +f_+(E_q)  \right] ,
  \label{eq:sg2}
\end{align}
and $E_q=\sqrt{\bm{q}^2+m^2}$.
The functions
$n(x)$ and $f_\pm(x)$ are the Bose-Einstein and the Fermi-Dirac
distribution functions,
$ n(x) = [ \exp(x/T) - 1 ]^{-1} $ and 
$ f_\pm(x) = [ \exp( (x \pm\mu)/T) + 1 ]^{-1}$, respectively.
For the evaluation of these equations, we first compute the imaginary
parts for which the $q_0$ integral is performed analytically
and the $q$ integral is carried out with the cutoff $\Lambda$,
and then evaluate the real parts with a relation \cite{Kitazawa:2005mp}
\begin{equation}
 {\rm Re}\Sigma_{\sigma(\pi)}^R(p_0)=-\frac{1}{\pi}{\rm P}
  \int_{-\Lambda}^{\Lambda}dp_0' \frac{{\rm Im}\Sigma_{\sigma(\pi)}^R(p_0')}{p_0-p_0'},
  \label{eq:qkk}
\end{equation}
where the cutoff $\Lambda$ is the same as that used in Sec.~\ref{sec:sigma}.

The retarded quark propagator for zero momentum reads
\begin{align}
  G^R( p_0 ) 
  &= \frac1{ (p_0 + i\eta +\mu) \gamma^0 - m - \Sigma^R (p_0) },
  \label{eq:G^R}
\end{align}
which is decomposed  into
\begin{equation}
  G^R(p_0) = G_+(p_0)\Lambda_+\gamma^0 + G_-(p_0)\Lambda_-\gamma^0
\end{equation}
using the projection operators $\Lambda_\pm = (1\pm\gamma_0)/2$.
Here $G_+(G_-)$ is the retarded quark (antiquark) propagator defined as
\begin{align}
  G_\pm ( p_0 ) 
  = \frac12 {\rm Tr} [ G^R \gamma^0 \Lambda_\pm ]
  = \frac1{ p_0 \mp m +  \mu - \Sigma^\pm ( p_0 ) } ,
  \label{eq:Gpm}
\end{align}
with $\Sigma_\pm(p_0)={\rm Tr}[\Sigma^R\Lambda_\pm\gamma^0]/2$.
The quasiquark $(\rho_+)$ and quasiantiquark $(\rho_-)$ spectral functions are then expressed
with $G_\pm$ as 
\begin{equation}
  \rho_\pm(p_0)=-\frac{1}{\pi}\textrm{Im}G_\pm ( p_0 ) .
  \label{spct}
\end{equation}
It is noted that $p_0=0$ denotes the Fermi level at $T=0$,
while $p_0+\mu=0$ is the Dirac level in our notation.

\section{Soft mode and mesonic modes}
\label{sec:softmode}

In this section,  we  summarize the behavior of the fluctuation modes 
at nonzero $T$ and $\mu$ as the basic ingredients for the analysis of 
the quark propagator.

\begin{figure}[h]
\hspace*{-3em}
\includegraphics[width=9.5cm]{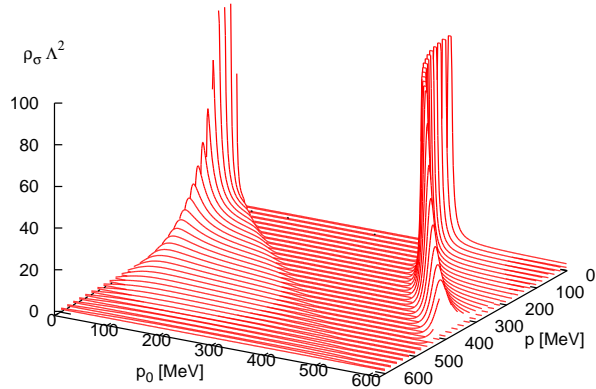}
\caption{
The spectral function $\rho_\sigma(\bm{p},p_0)$
at $T=48$ MeV and $\mu\simeq 328$ MeV.}
\label{fig:rhos}
\end{figure}

Let us first focus on the scalar channel.
Figure~\ref{fig:rhos}  shows the spectral function $\rho_\sigma(\bm{p}, p_0)$
at $T= 48$ MeV and $\mu\simeq328$ MeV, 
which is on the ridge of the chiral susceptibility and very close to the CP.
We see that there is not only a peak in the timelike region corresponding
to the sigma mode \cite{Kunihiro:2007bx} but also 
 a sharp peak in the spacelike region
around $p\approx0$ and $p_0\approx 0$,
the latter of which is precisely the soft mode associated with the CP.
The small width of the soft mode 
comes from the Landau damping that is effective in the spacelike region.
The propagator $D^R_\sigma(\bm{p},p_0)$ in the scalar channel has 
a pole associated with the soft mode:
The pole moves toward and eventually reaches the origin
in the complex energy plane as the system approaches the CP; i.e.,
both the mass and the width of the soft mode vanish at the CP.
The nature of the soft mode 
is actually the collective particle-hole excitation familiar in many-body physics, 
which is possible owing to the 
presence of the density (or the Fermi sphere).

\begin{figure}[t]
\includegraphics[width=8cm]{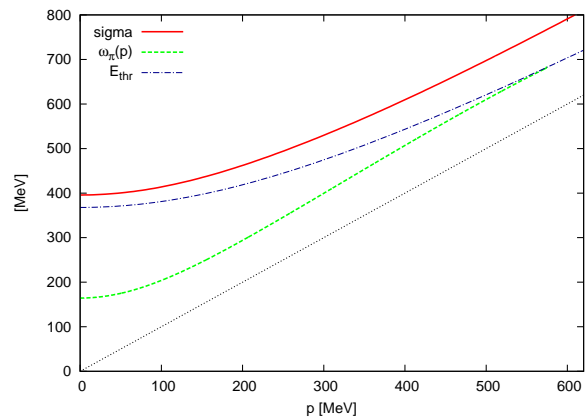}
\caption{The peak position of the sigma mode,
the dispersion relation of the stable pionic mode, $\omega_\pi(p)$,
and the threshold energy, $E_{\rm thr}=\sqrt{p^2+4m^2}$,
at $T=48$ MeV and $\mu\simeq 328$ MeV.
}
\label{fig:peaks}
\end{figure}

In contrast, the sigma mode in the timelike region
has a nonzero mass and the width even at the CP, the latter of which
gradually broadens as the momentum increases.
In Fig.~\ref{fig:peaks}, the peak position of the sigma mode,
which is approximately regarded as the mass of the sigma mode, is plotted
at the same $T$ and $\mu$ as those in Fig.~\ref{fig:rhos}.

Next we turn to the pseudoscalar channel.
As $T$ and/or $\mu$ are raised, the constituent quark mass $m$ becomes smaller
while the rest mass of the pionic mode $\omega_\pi(0)$ becomes larger.
Then,  the pionic mode acquires a width since 
the decay into a quark and an antiquark pair becomes possible.
We call the temperature at which the rest pionic mode becomes unstable as 
the pion zero-binding temperature $T_{\rm ZB}$, which
is determined by
\begin{equation}
  [{\rm Re}D_\pi^R(\bm{0}, 2m)]^{-1}_{T=T_{\rm ZB}}=0 ,
  \label{pidr}
\end{equation}
for a given $\mu$.
In Fig.~\ref{fig:phased}, 
$T_{\rm ZB}$
as a function of $\mu$ is plotted with the dotted line. 
We see that $T_{\rm ZB}$ is always higher than 
the dashed line representing the temperature at which 
the chiral susceptibility has the maximum for a given $\mu$.
It is also noted that 
the CP is located below $T_{\rm ZB}$ in our analysis, 
which means that the pionic modes are stable at the CP.
As $\mu$ goes high, $T_{\rm ZB}$ decreases and eventually merges into the critical line of the 
first-order transition at $\mu\gtrsim 340$ MeV.

The decay of a pionic mode into a quark and an antiquark 
pair is also possible when the dispersion relation $\omega_\pi(p)$ becomes larger than 
$E_{\rm thr}(p)$ even for $T<T_{\rm ZB}$,
which implies that the pionic modes with a large velocity relative to the medium
can be unstable owing to the medium effect.
In Fig.~\ref{fig:peaks}, the dispersion relation $\omega_\pi(p)$ 
is also plotted; one sees that
$\omega_\pi(p)$ becomes larger than 
$E_{\rm thr}(p)$ at $p\simeq 575$MeV.
This is due to the fact that the dispersion relation of the pionic modes
is not the hyperbolic form, 
$\omega_\pi(p)\neq \sqrt{p^2+\omega_\pi^2(0)}$.
In a previous study \cite{KKN},  the present authors showed that 
such an anomalous dispersion relation leads to a divergence of the joint density of
states of the pionic modes and a quark, 
and gives rise to a van Hove singularity in the quark self energy.
Although this is also 
the case near the CP,
it will be found that the singularity is negligibly weak.


\section{Numerical results and discussions}
\label{sec:num}

In this section,
we present the numerical results for the quark spectrum around the CP.

\begin{figure*}[t]
\hspace{-2em}
\includegraphics[width=16.5cm]{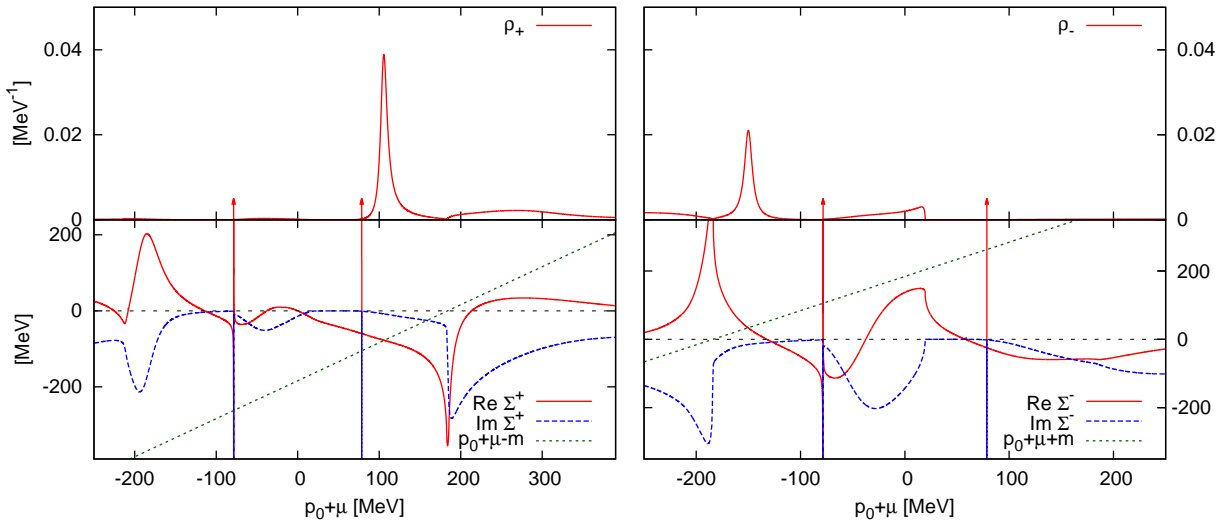}
\caption{
The upper left (right) panel is the quark (antiquark) spectral function $\rho_+(p_0)
(\rho_-(p_0))$
for $T=48$ MeV and $\mu\simeq 328$ MeV.
Peaks with very small spectral weights ($Z\lesssim 10^{-2}$)
are shown in the arrows.
The lower panels represent the real and imaginary 
parts of the corresponding quark self energies $\Sigma^\pm(p_0)$.
The dashed lines denotes $p_0+\mu\mp m$ with $m$ being the constituent
quark mass.}
\label{fig:tp048ap}
\end{figure*}

\subsection{Near the critical point} \label{nearCP}

We first see the quark spectral function $\rho_+(p_0)$ 
in the vicinity of the CP.
The upper left panel of Fig.~\ref{fig:tp048ap} shows
the quark spectral function $\rho_+(p_0)$ at $T=48$ MeV and $\mu\simeq 328$ MeV,
which is very close to the CP and on the ridge of the chiral susceptibility.
We note that the quark spectral function in the MFA is given by
$\rho_+(p_0)=\delta(p_0+\mu-m)$ with $m\simeq184$ MeV being the corresponding constituent quark
mass in the MFA.
Figure~\ref{fig:tp048ap} shows that 
the quark spectrum is largely modified from the one in the MFA;
a peak given by the MFA at $p_0=m-\mu$ is totally absent. Moreover,
one finds a prominent peak at an unexpectedly smaller energy than $m$, i.e.,
 at $p_0+\mu \simeq  105$ MeV; we note that the collective excitation corresponding 
to the peak at a positive energy has a positive  quark number.

\begin{figure}[t]
\includegraphics[width=3cm]{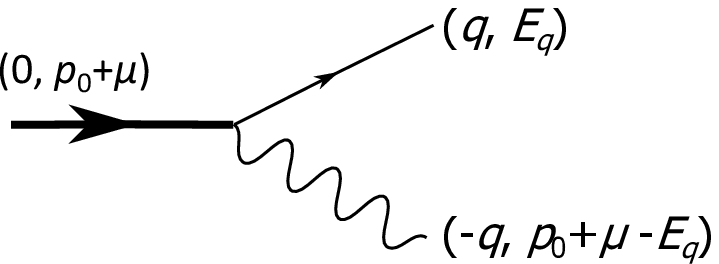}
\hspace{1em}
\includegraphics[width=3cm]{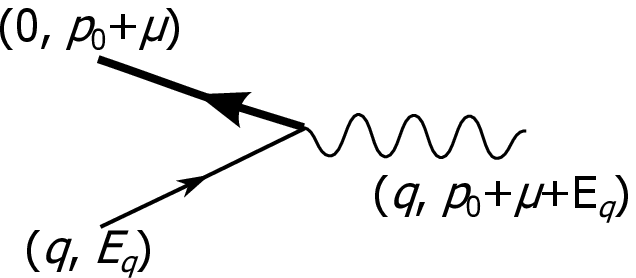}
\caption{The scattering processes that form the peaks
in Im$\Sigma_+$ at $p_0+\mu\simeq 190$MeV (left) and
Im$\Sigma_-$ at $p_0+\mu\simeq-190$ MeV (right).
The thick solid line represents the quasiquark with $(\bm{0},p_0+\mu)$,
the thin  solid line the on-shell free quark, and the wavy line
the soft mode.
The inverse processes are also possible.
}
\label{fig:fd1}
\end{figure}

To figure out the origin of this anomalous peak, 
let us make a detailed 
analysis of the corresponding quark self energy.
We show the real and imaginary parts of $\Sigma^+(p_0)$ in the lower left panel of 
Fig.~\ref{fig:tp048ap}.
We see that there is a broad peak in Im$\Sigma_+(p_0)$ around $p_0+\mu\simeq190$ MeV,
implying 
that the single quark state with this energy 
decays into other states.
Our numerical result with the decomposition 
Eqs.~(\ref{eq:sg})-(\ref{eq:sg2}) shows that 
this peak of Im$\Sigma^+(p_0)$ comes from the decay process of
a quark into the soft mode shown in the left panel of Fig.~\ref{fig:fd1}.
Because the strength of the soft mode is enhanced when both the energy and the momentum
are close to zero in the spacelike region,
this process
is enhanced at $q\approx 0$
and $p_0+\mu-E_q\approx 0$.
In addition, from Eq.~(\ref{eq:sg1}), one sees that the Bose-Einstein distribution 
function in the $q$ integral in Im$\Sigma^+(p_0)$ 
also acts to enhance the strength of the decay process around 
$p_0+\mu-E_q= 0$.

\begin{figure*}[t]
\hspace{-2em}
\includegraphics[width=16.5cm]{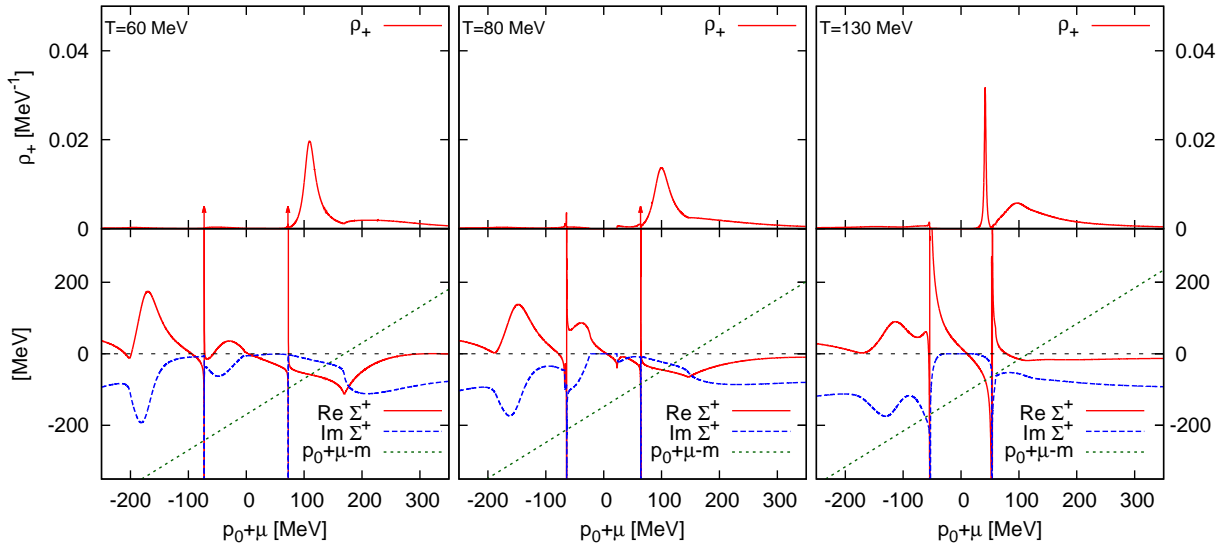}
\caption{
The same as Fig.~\ref{fig:tp048ap} but at $T=60$ MeV and
$\mu\simeq 320$ MeV (left), 
$T=80$ MeV and $\mu\simeq304$ MeV (middle), and
$T=130$ MeV and $\mu\simeq258$ MeV (right). 
These are on the ridge of the chiral susceptibility.}
\label{fig:3spc}
\end{figure*}

The existence of a large peak in Im$\Sigma^+(p_0)$ causes a steep rise in
Re$\Sigma^+(p_0)$ around it owing to the analytic property
Eq.~(\ref{eq:qkk})
 which in turn 
gives rise to a zero in the real part 
of the inverse propagator 
 for the quark (antiquark),
\begin{equation}
  p_0+\mu\mp m-{\rm Re}\Sigma^\pm(p_0)=0 .
  \label{eq:qdisp}
\end{equation}
When Im$\Sigma^\pm(p_0)$ is small at $p_0$ satisfying Eq.~(\ref{eq:qdisp}), 
the quark propagator has a peak there \cite{Kitazawa:2005mp}.
The solutions of Eq.~(\ref{eq:qdisp}) 
are graphically determined by points of intersection
 between ${\rm Re}\Sigma^+(p_0)$ and a line $p_0+\mu-m$ 
that is shown by a dashed green line
in the lower left panel of Fig.~\ref{fig:tp048ap}.
One sees that Re$\Sigma^+(p_0)$ shows a peaklike structure
around the peak in Im$\Sigma^+(p_0)$ at $p_0+\mu\simeq190$ MeV.
Consequently, 
this bending behavior of Re$\Sigma^+(p_0)$ causes the shift of the position of the 
quasipole from that of the pole in the MFA.
For $190\lesssim p_0+\mu\lesssim 300$ MeV, 
there is no clear peak
but a bumplike structure in $\rho_+(p_0)$, 
because
$|\textrm{Im}\Sigma^+(p_0)|$ keeps large values and there is no quasipole.

We remark that the arrows at $p_0+\mu\simeq \pm80$ MeV 
in the upper left panel of  Fig.~\ref{fig:tp048ap}
indicate two sharp peaks in  $\rho_+(p_0)$ with a vanishingly narrow 
width.
The spectral weights of these peaks, however, are negligibly small.
In fact, by defining the spectral weight of the peak by
\begin{equation}
  Z = \int_\Delta dp_0 \rho_\pm(p_0),
  \label{eq:QuasiRes}
\end{equation}
where $\Delta$ is a range of $p_0$ that well covers the peak, 
one gets $Z\lesssim 10^{-2}$ for these peaks, while
the sum of the spectral weight is $\int dp_0\rho_+(p_0)=1$
in our definition.
Despite the sharpness of the peak structure, therefore, 
these peaks do not have physical significance.

The emergence of these peaks is understood by the van Hove
singularity triggered by the nonhyperbolic dispersion 
relation of the stable pionic modes $\omega_\pi(p)$ \cite{KKN}.
As discussed in Sec.~\ref{sec:softmode}, $\omega_\pi(p)$ has 
a nonhyperbolic form near the CP owing to the violation of 
Lorentz symmetry in medium.
In Ref.~\cite{KKN}, it is discussed that such a distorted 
dispersion relation can give rise to zeros of the relative 
velocity between the quark and the bound pion, 
and hence the divergences of the joint density of states 
$[d\{E_q-\omega_\pi(q)\}/dq]^{-1}$ at some energies.
Reflecting these divergences, $\Sigma^+(p_0)$ and $\rho_+(p_0)$ 
also diverge at these energies as the van Hove singularity.

In Ref.~\cite{KKN}, we have shown that the van Hove singularity 
can significantly alter the spectral properties of quarks near the 
pseudocritical temperature at $\mu=0$.
While the same singularity manifests itself even near the CP, 
the strength of the singularity is weak.
Therefore, 
the singularity near the CP would easily be blurred by 
the effects neglected in the present analysis, such as 
the contribution of higher-order terms of the quark self energy,
in contrast to the case discussed in Ref.~\cite{KKN}.
We thus do not discuss this structure in more detail in the present study.

The spectral function for the antiquark sector $\rho_-(p_0)$
for the same $T$ and $\mu$ 
is shown in the upper right panel of
Fig. \ref{fig:tp048ap}.
One sees a peak at $p_0+\mu\simeq -145$ MeV whose position is
shifted from that in the MFA, $-m=-184$ MeV.
To identify the origin of the shift, we show
 Im$\Sigma^-(p_0)$ in the lower right panel of Fig.~\ref{fig:tp048ap}.
One finds a peak in Im$\Sigma^-(p_0)$ at 
$p_0+\mu\simeq-190$ MeV.
Our numerical analysis shows that this peak 
corresponds to 
a scattering process of an antiquark and a quark
into the soft mode as denoted in the right panel of Fig.~\ref{fig:fd1}.
Then, similarly to the previous discussion on $\rho_+(p_0)$,
the peak distorts Re$\Sigma^-(p_0)$
and gives rise to a new quasipole at $p_0+\mu\simeq-190$ MeV.
In this way, the antiquark sector also receives considerable
modification from the soft mode near the CP.
In $\rho_-(p_0)$, we also find peaks with vanishingly narrow
widths as denoted by arrows in the upper right panel.
The spectral strengths of these quasipoles are both negligible, $Z\lesssim 10^{-2}$.

\subsection{The case away from the critical point} \label{apartCP}

 In this section, we examine  how the spectral properties of 
the quark are
affected by 
the soft mode associated with the CP
when the system is away from the CP; we
show $\rho_+(p_0)$ for some  combination of $(T,\, \mu)$ with $T>T_c$ and $\mu<\mu_c$.
Although we only show $\rho_+(p_0)$,
the spectral properties of the antiquark is qualitatively the same
as those of the quark.

We first see the quark spectrum by varying $T$ and $\mu$ 
along the ridge of the chiral susceptibility.
The quark spectral functions $\rho_+(p_0)$ and the self energies
$\Sigma^+(p_0)$ 
on the ridge
are shown in Fig.~\ref{fig:3spc}
for $T=60$, $80$, and $130$ MeV; 
the corresponding chemical potentials are 
$\mu\simeq320$, $304$, and $258$ MeV, respectively.
Behaviors of $\rho_+(p_0)$ and $\Sigma^+(p_0)$ for $T=60$ MeV
shown in the far-left panels are similar to those plotted 
in Fig.~\ref{fig:tp048ap},
whereas clear peak structures in these functions,
i.e., the peaks in $\rho_+(p_0)$ at $p_0+\mu\simeq105$ MeV  and 
Im$\Sigma^+(p_0)$ at $p_0+\mu\simeq190$ MeV,
observed near the CP are slightly obscured for this $T$.
This result shows that the soft mode associated with the CP can affect 
the quark spectrum even apart from the CP
on the ridge of the chiral susceptibility.
As $T$ is raised further, the effect of the soft mode gradually 
ceases to exist as shown in the middle panels of Fig.~\ref{fig:3spc}.
In the lower middle panel one finds that the bump structure
at $p_0+\mu \simeq 190$ MeV almost disappears
for $T=80$ MeV. 
Accordingly, the energy shift of the position of the 
quasiquark peak compared with the MFA value 
becomes smaller as shown in the upper middle panel.
The lower right panel shows that the bump structure
in Im$\Sigma^+(p_0)$ disappears almost completely for $T=130$ MeV.

In the upper right panel of Fig.~\ref{fig:3spc}, one finds that 
$\rho_+(p_0)$ again forms a sharp peak at $p_0+\mu$ significantly
smaller than the constituent quark mass $m$, although the effect
of the soft mode associated with the CP is suppressed.
From the behavior of $\Sigma^+(p_0)$ shown in the lower right panel, 
one finds that this modification comes from the singular behavior 
of $\Sigma^+(p_0)$ around $p_0+\mu \simeq 50$ MeV.
As discussed in Ref.~\cite{KKN}, 
these singularities are the van Hove singularity in the 
scattering process of quarks and pionic modes
induced by the nonhyperbolic dispersion relation of the
pionic modes.
This result shows that the effect of the pionic modes
becomes more prominent as $\mu$ is lowered
instead of the soft mode associated with the CP.

Off the ridge of the chiral susceptibility, the strength of the soft mode rapidly decreases as
$T$ and/or $\mu$ are 
away from the CP.
The strength of the peak in the quark spectrum originated from the scattering of
the soft mode accordingly
decreases rapidly.
For example, when $T$ is raised from the CP with fixed $\mu=\mu_c$,
the distortion of the quark spectrum due to the soft mode is 
insignificant already at $T\simeq 1.05T_c$.

\section{Brief summary and concluding remarks}
\label{sec:conc}

In this paper,
we have investigated the quark spectrum
at nonzero temperature and density near the CP of  
the chiral transition in an effective model.
The soft mode associated with the second-order transition at the CP is 
the scalar-density fluctuation in the spacelike region
\cite{Fujii:2003bz}.
In the present study, we have investigated the effect of 
this soft mode on the quark spectrum for the first time.
We have shown that
the soft mode  strongly couples to a quark and an antiquark 
at vanishing momentum near the CP.
As a result of this coupling,
a quasiquark peak is created 
at the energy significantly lower than that in the mean-field approximation.
It is also found that the strong modification of the quark spectrum
due to the soft mode is observed over a wide range of $T$ and $\mu$
on the ridge of the chiral susceptibility. 
The effect of the soft mode, however, is suppressed rapidly 
off the ridge.
In our previous study \cite{KKN}, we found that 
the quark spectrum near the pseudocritical temperature at $\mu=0$ 
is strongly modified by the van Hove singularity \cite{KKN}.
Although such a singularity is observed even for nonzero $\mu$, 
the effect of the singularity is negligibly weak near the CP.

In the present study, we have limited our attention to 
the quark spectral function at zero momentum,
since the soft mode influences the quark spectrum at vanishing momentum
most strongly.
On the other hand,
excitation modes near the Fermi momentum are
relevant degrees of freedom in high density but low-temperature systems.
The study of modes near the Fermi surface is left as a future work.
As mentioned in Sec.~\ref{sec:sigma}, 
a complete analysis of the 
quark spectrum near the CP will require us to 
incorporate all the hydrodynamic modes including the number-density fluctuation (phonon) and entropy
fluctuations.
The coupling to the density fluctuations may be
taken into account by the inclusion of the vector interaction \cite{Kunihiro:1991qu} within this model.
There are some attempts to incorporate the 
fermion-hydrodynamic modes coupling 
in the analysis of the phase structure and transport properties of the many body systems
\cite{Kamikado:2012cp,Shen:2013kxa}.
We hope that 
we can report on such an  analysis that extensively takes care of
the coupling of the hydrodynamic modes to the quark near the CP elsewhere.

This work was supported by JSPS KAKENHI Grant No. 25800148,
No. 24340054, and No. 24540271.
T.K. was partially supported 
 by  the Core Stage Back UP program 
in Kyoto Univeristy, and by the Yukawa
International Program for Quark-hadron Sciences.



\begin{thebibliography}{99}

\bibitem{Bernard:2004je} 
  C.~Bernard {\it et al.}  [MILC Collaboration],
  Phys.\ Rev.\ D {\bf 71}, 034504 (2005)
  [hep-lat/0405029].

\bibitem{Cheng:2006qk} 
  M.~Cheng {\it et al.},
  Phys.\ Rev.\ D {\bf 74}, 054507 (2006)
  [hep-lat/0608013].

\bibitem{Aoki:2006we} 
  Y.~Aoki, G.~Endrodi, Z.~Fodor, S.~D.~Katz and K.~K.~Szabo,
  Nature {\bf 443}, 675 (2006)
  [hep-lat/0611014].

\bibitem{Ejiri:2009ac} 
  S.~Ejiri {\it et al.},
  Phys.\ Rev.\ D {\bf 80}, 094505 (2009)
  [arXiv:0909.5122 [hep-lat]].

\bibitem{Fodor:2004nz} 
  Z.~Fodor and S.~D.~Katz,
  JHEP {\bf 0404}, 050 (2004)
  [hep-lat/0402006].

\bibitem{Gavai:2004sd} 
  R.~V.~Gavai and S.~Gupta,
  Phys.\ Rev.\ D {\bf 71}, 114014 (2005)
  [hep-lat/0412035].

\bibitem{Ejiri:2013lia} 
  S.~Ejiri,
  Eur.\ Phys.\ J.\ A {\bf 49}, 86 (2013)
  [arXiv:1306.0295 [hep-lat]].

\bibitem{Aarts:2013uxa} 
  G.~Aarts, L.~Bongiovanni, E.~Seiler, D.~Sexty and I.~-O.~Stamatescu,
  Eur.\ Phys.\ J.\ A {\bf 49}, 89 (2013)
  [arXiv:1303.6425 [hep-lat]].

\bibitem{Philipsen:2011zx} 
  O.~Philipsen,
  Acta Phys.\ Polon.\ Supp.\  {\bf 5}, 825 (2012)
  [arXiv:1111.5370 [hep-ph]].

\bibitem{Klevansky:1992qe} 
  S.~P.~Klevansky,
  Rev.\ Mod.\ Phys.\  {\bf 64}, 649 (1992).

\bibitem{Hatsuda:1994pi}
  T.~Hatsuda and T.~Kunihiro,
  Phys.\ Rept.\  {\bf 247}, 221 (1994)
  [arXiv:hep-ph/9401310].

\bibitem{Buballa:2003qv} 
  M.~Buballa,
  Phys.\ Rept.\  {\bf 407}, 205 (2005)
  [hep-ph/0402234].

\bibitem{Asakawa:1989bq}
  M.~Asakawa and K.~Yazaki,
  Nucl.\ Phys.\  A {\bf 504}, 668 (1989).

\bibitem{Kunihiro:1991qu}
  T.~Kunihiro,
  Phys.\ Lett.\ B {\bf 271} (1991) 395.

\bibitem{Kitazawa:2002bc} 
  M.~Kitazawa, T.~Koide, T.~Kunihiro and Y.~Nemoto,
  Prog.\ Theor.\ Phys.\  {\bf 108}, 929 (2002)
  [hep-ph/0207255, hep-ph/0307278].

\bibitem{Zhang:2008wx} 
  Z.~Zhang, K.~Fukushima and T.~Kunihiro,
  Phys.\ Rev.\ D {\bf 79}, 014004 (2009)
  [arXiv:0808.0927 [hep-ph]];
%
  Z.~Zhang and T.~Kunihiro,
  Phys.\ Rev.\ D {\bf 80}, 014015 (2009)
  [arXiv:0904.1062 [hep-ph]].

\bibitem{Hatsuda:2006ps} 
  T.~Hatsuda, M.~Tachibana, N.~Yamamoto and G.~Baym,
  Phys.\ Rev.\ Lett.\  {\bf 97}, 122001 (2006)
  [hep-ph/0605018];
%
  N.~Yamamoto, M.~Tachibana, T.~Hatsuda and G.~Baym,
  Phys.\ Rev.\ D {\bf 76}, 074001 (2007)
  [arXiv:0704.2654 [hep-ph]].

\bibitem{deForcrand:2006pv} 
  P.~de Forcrand and O.~Philipsen,
  JHEP {\bf 0701}, 077 (2007)
  [hep-lat/0607017];
%
  JHEP {\bf 0811}, 012 (2008)
  [arXiv:0808.1096 [hep-lat]];
%
  O.~Philipsen,
  Prog.\ Theor.\ Phys.\ Suppl.\  {\bf 174}, 206 (2008)
  [arXiv:0808.0672 [hep-ph]];
%
  K.~Fukushima,
  Phys.\ Rev.\ D {\bf 78}, 114019 (2008)
  [arXiv:0809.3080 [hep-ph]];
%
  J.~W.~Chen, K.~Fukushima, H.~Kohyama, K.~Ohnishi and U.~Raha,
  Phys.\ Rev.\ D {\bf 80}, 054012 (2009)
  [arXiv:0901.2407 [hep-ph]].

\bibitem{Basler:2010xy}
  H.~Basler and M.~Buballa,
   Phys.\ Rev.\  D {\bf 82}, 094004 (2010)
   arXiv:1007.5198 [hep-ph].

\bibitem{Zhang:2011xi}
  Z.~Zhang and T.~Kunihiro,
  Phys.\ Rev.\ D {\bf 83}, 114003 (2011)
  [arXiv:1102.3263 [hep-ph]].

\bibitem{Nakano:2004cd}
  E.~Nakano and T.~Tatsumi,
  Phys.\ Rev.\ D {\bf 71}, 114006 (2005)
  [hep-ph/0411350].

\bibitem{Nickel:2009ke}
  D.~Nickel,
  Phys.\ Rev.\ Lett.\  {\bf 103}, 072301 (2009)
  [arXiv:0902.1778 [hep-ph]];
%
  Phys.\ Rev.\ D {\bf 80}, 074025 (2009)
  [arXiv:0906.5295 [hep-ph]].

\bibitem{Aggarwal:2010wy} 
  M.~M.~Aggarwal {\it et al.}  [STAR Collaboration],
  Phys.\ Rev.\ Lett.\  {\bf 105}, 022302 (2010)
  [arXiv:1004.4959 [nucl-ex]];
%
  L.~Adamczyk {\it et al.}  [STAR Collaboration],
  Phys.\ Rev.\ Lett.\  {\bf 112}, no. 3, 032302 (2014)
  [arXiv:1309.5681 [nucl-ex]].

\bibitem{Stephanov:1998dy} 
  M.~A.~Stephanov, K.~Rajagopal and E.~V.~Shuryak,
  Phys.\ Rev.\ Lett.\  {\bf 81}, 4816 (1998)
  [hep-ph/9806219];
%
  Phys.\ Rev.\ D {\bf 60}, 114028 (1999)
  [hep-ph/9903292];
%
  M.~Asakawa, S.~Ejiri and M.~Kitazawa,
  Phys.\ Rev.\ Lett.\  {\bf 103}, 262301 (2009)
  [arXiv:0904.2089 [nucl-th]];
%
  M.~Kitazawa, M.~Asakawa and H.~Ono,
  Phys.\ Lett.\ B {\bf 728}, 386 (2014)
  [arXiv:1307.2978].

\bibitem{Fujii:2003bz}
  H.~Fujii,
  Phys.\ Rev.\  D {\bf 67}, 094018 (2003)
  [arXiv:hep-ph/0302167];
%
  H.~Fujii and M.~Ohtani,
  Phys.\ Rev.\  D {\bf 70}, 014016 (2004)
  [arXiv:hep-ph/0402263].

\bibitem{Son:2004iv} 
  D.~T.~Son and M.~A.~Stephanov,
  Phys.\ Rev.\ D {\bf 70}, 056001 (2004)
  [hep-ph/0401052].

\bibitem{Hatsuda:1984jm} 
  T.~Hatsuda and T.~Kunihiro,
  Phys.\ Lett.\ B {\bf 145}, 7 (1984);
%
  Phys.\ Rev.\ Lett.\  {\bf 55}, 158 (1985).

\bibitem{Hohenberg:1977ym}
  P.~C.~Hohenberg and B.~I.~Halperin,
  Rev.\ Mod.\ Phys.\  {\bf 49}, 435 (1977).

\bibitem{Nambu:1961tp}
Y.~Nambu and G.~Jona-Lasinio,
  Phys.\ Rev.\  {\bf 122}, 345 (1961);
%
  Phys.\ Rev.\  {\bf 124}, 246 (1961).

\bibitem{Frenkel:1989br} 
  J.~Frenkel and J.~C.~Taylor,
  Nucl.\ Phys.\ B {\bf 334}, 199 (1990);
%
  E.~Braaten and R.~D.~Pisarski,
  Nucl.\ Phys.\ B {\bf 339}, 310 (1990);
%
 H.~A.~Weldon,
  Phys.\ Rev.\ D {\bf 26}, 2789 (1982);
%
  Phys.\ Rev.\ D {\bf 40}, 2410 (1989).

\bibitem{LeB}
  M.~Le Bellac, {\it Thermal Field Theory}, 
 (Cambridge University Press, Cambridge, 1996).

\bibitem{Kitazawa:2005mp}
  M.~Kitazawa, T.~Kunihiro and Y.~Nemoto,
  Phys.\ Lett.\  B {\bf 633}, 269 (2006)
  [arXiv:hep-ph/0510167].

\bibitem{Kitazawa:2006zi}
  M.~Kitazawa, T.~Kunihiro and Y.~Nemoto,
  Prog.\ Theor.\ Phys.\  {\bf 117}, 103 (2007)
  [arXiv:hep-ph/0609164].

\bibitem{Satow:2010ia} 
  D.~Satow, Y.~Hidaka and T.~Kunihiro,
  Phys.\ Rev.\ D {\bf 83}, 045017 (2011)
  [arXiv:1011.6452 [hep-ph]].

\bibitem{Harada:2007gg} 
  M.~Harada, Y.~Nemoto and S.~Yoshimoto,
  Prog.\ Theor.\ Phys.\  {\bf 119}, 117 (2008)
  [arXiv:0708.3351 [hep-ph]].

\bibitem{Harada:2008vk} 
  M.~Harada and Y.~Nemoto,
  Phys.\ Rev.\ D {\bf 78}, 014004 (2008)
  [arXiv:0803.3257 [hep-ph]].

\bibitem{Qin:2010pc} 
  S.~-x.~Qin, L.~Chang, Y.~-x.~Liu and C.~D.~Roberts,
  Phys.\ Rev.\ D {\bf 84}, 014017 (2011)
  [arXiv:1010.4231 [nucl-th]];
%
  A.~Bashir, L.~Chang, I.~C.~Cloet, B.~El-Bennich, Y.~-X.~Liu, C.~D.~Roberts and P.~C.~Tandy,
  Commun.\ Theor.\ Phys.\  {\bf 58}, 79 (2012)
  [arXiv:1201.3366 [nucl-th]].

\bibitem{Nakkagawa:2011ci} 
  H.~Nakkagawa, H.~Yokota and K.~Yoshida,
  Phys.\ Rev.\ D {\bf 85}, 031902 (2012)
  [arXiv:1111.0117 [hep-ph]];
%
  Phys.\ Rev.\ D {\bf 86}, 096007 (2012)
  [arXiv:1208.6386].

\bibitem{Qin:2013ufa} 
  S.~-x.~Qin and D.~H.~Rischke,
  Phys.\ Rev.\ D {\bf 88}, 056007 (2013)
  [arXiv:1304.6547 [nucl-th]].

\bibitem{Hidaka:2011rz} 
  Y.~Hidaka, D.~Satow and T.~Kunihiro,
  Nucl.\ Phys.\ A {\bf 876}, 93 (2012)
  [arXiv:1111.5015 [hep-ph]].

\bibitem{Miura:2013fxa}
  K.~Miura, Y.~Hidaka, D.~Satow and T.~Kunihiro,
  Phys.\ Rev.\ D {\bf 88}, 065024 (2013)
  [arXiv:1306.1701 [hep-ph]].

\bibitem{Kitazawa:2007ep} 
  M.~Kitazawa, T.~Kunihiro, K.~Mitsutani and Y.~Nemoto,
  Phys.\ Rev.\ D {\bf 77}, 045034 (2008)
  [arXiv:0710.5809 [hep-ph]].

\bibitem{KKN}
  M.~Kitazawa, T.~Kunihiro and Y.~Nemoto,
  Phys.\ Rev.\  D {\bf 89}, 056002 (2014) 
  [arXiv:1312.3022 [hep-ph]].

\bibitem{HL}
  L.~Hedin and S.~Lundqvist, Solid State Phys. {\bf 23}, 1 (1970).

\bibitem{Kamikado:2012cp} 
  K.~Kamikado, T.~Kunihiro, K.~Morita and A.~Ohnishi,
  PTEP {\bf 2013}, 053D01 (2013)
  [arXiv:1210.8347 [hep-ph]].

\bibitem{Shen:2013kxa} 
  G.~Shen and S.~Reddy,
  arXiv:1311.6096 [nucl-th].

\bibitem{Kunihiro:2007bx}
  T.~Kunihiro, M.~Kitazawa and Y.~Nemoto,
  Proc. Sci. {\bf CPOD07}  (2007) 041, 
  arXiv:0711.4429.

\end{thebibliography}
\end{document}